\documentclass[%
reprint,
superscriptaddress,
amsmath,amssymb,
aps,
prx,
tightenlines,
longbibliography
]{revtex4-1}

\usepackage{psfrag,graphicx,epsfig,color}
\usepackage{dcolumn}
\usepackage{bm}
\usepackage{natbib}
\usepackage[usenames,dvipsnames,svgnames,table]{xcolor}
\usepackage{subfigure}
\usepackage{rotating}
\usepackage{float}

\def\epsav {{\langle \epsilon \rangle}}
\def\re    {{R_\lambda}}

\begin{document}

\title{Dissipation range of the energy spectrum in high Reynolds number turbulence}

\author{Dhawal Buaria }
\email[]{dhawal.buaria@nyu.edu}
\affiliation{Tandon School of Engineering, New York  University, New York, NY 11201, USA}
\affiliation{Max Planck Institute for Dynamics and Self-Organization, 37077 G\"ottingen, Germany}

\author{Katepalli R. Sreenivasan}
\affiliation{Tandon School of Engineering, New York  University, New York, NY 11201, USA}
\affiliation{Department of Physics and the Courant Institute of Mathematical Sciences,
New York University, New York, NY 10012, USA}

\date{\today}


\begin{abstract}

We seek to understand the kinetic energy spectrum in the dissipation range of fully developed turbulence. The data are obtained by direct numerical simulations (DNS) of forced Navier-Stokes equations in a periodic domain, for Taylor-scale Reynolds numbers up to $\re=650$, with excellent small-scale resolution of $k_{max}\eta \approx 6$,  
and additionally at $\re=1300$ with $k_{max}\eta\approx3$, where $k_{max}$ is the maximum resolved wavenumber and $\eta$ is the Kolmogorov length scale. 
We find that for a limited range of wavenumbers $k$ past the bottleneck, in the range $0.15 \lesssim k\eta \lesssim0.5$, the spectra for all $\re$ display a universal stretched exponential behavior of the form $\exp(-k^{2/3})$, in rough accordance with recent theoretical predictions. In contrast, the stretched exponential fit does not possess a unique exponent in the near dissipation range $1 \lesssim k\eta \lesssim 4$, but one that decreases with increasing $\re$. This region serves as 
the intermediate dissipation range between the stretched exponential behavior and the far dissipation range $k\eta \gg 1$ where analytical arguments as well as DNS data with superfine resolution (S. Khurshid et al., Phys.~Rev.~Fluids 3, 082601, 2018) suggest a simple $\exp(-k\eta)$ dependence. We briefly discuss our results in connection to the multifractal model.


\end{abstract}

\maketitle

\onecolumngrid

\paragraph*{Introduction:} Turbulent fluctuations in fluid flows span a wide range of scales and are often characterized by the energy spectrum $E(k)$, where $k$ is 
wavenumber i.e., the norm of the wave vector, whose inverse measures the scale size in real space \cite{Frisch95,popebook}. The integral of $E(k)$ over all $k$ gives the average kinetic energy of turbulence. The pioneering work of Kolmogorov \cite{k41}, K41 henceforth, theorized that the small scales are universal at sufficiently high Reynolds numbers, depending solely on the viscosity $\nu$ and the mean dissipation rate $\langle \epsilon \rangle$. In addition, at an intermediate range of scales, the so-called inertial range, the dependence on $\nu$ vanishes as well. These considerations imply that in the range of scales much smaller than the energy injection scale, the energy spectrum can be written as $E(k) \sim \epsav^{2/3} k^{-5/3} f(k\eta)$, where $\eta=(\nu^3/\epsav)^{1/4}$ is the Kolmogorov length scale and $f$ is some universal function of $k\eta$, tending to a constant in the inertial range. The energy spectrum has been extensively studied by numerous researchers, and the $k^{-5/3}$ prediction (with some small intermittency correction) seems to have received substantial validation \cite{sv94,sreeni95,fritts03,Ishihara09}. However, the functional form of $f$ and its universality in the dissipation range, are still not properly understood.

Many attempts have been made over the years to characterize $f$ using both experiments and direct numerical simulations (DNS) \cite{sreenivasan1985a, she1993a, chen1993a, martinez1997a, ishihara2005a, schum07sub, canet17, kds18, debue18}, all of which suggest the following general form 
\begin{align}
E(k\eta) \simeq (k\eta)^\alpha \exp \left[ -\beta (k\eta)^\gamma \right] \ .
\label{eq:gamma}
\end{align}
However, experiments have seldom resolved the range beyond $k\eta\approx1$ \cite{sreenivasan1985a,sv94}, and DNS have been either restricted to low $\re$ \cite{chen1993a,she1993a,martinez1997a,schum07sub} or achieved high $\re$ by sacrificing small-scale resolution \cite{ishihara2005a}. Consequently, there has been no clarity regarding the values of the coefficients in Eq.~\eqref{eq:gamma}, especially the exponent $\gamma$. The direct interaction approximation \cite{kraichnan1959a} and other ideas \cite{frisch1981a,sreenivasan1985a,foias1990a,sirovich1994a}, predict a pure exponential, i.e., $\gamma=1$ for large wavenumbers or very small scales regularized by viscosity. While this prediction was found to hold at low $\re$ \cite{chen1993a,martinez1997a,schum07sub}, it could not adequately describe data at higher $\re$ and often led to conflicting and ad hoc fits \cite{sreenivasan1985a,smith91,manley1992a,sv94}.

The above issues were addressed in a recent study \cite{kds18} by means of a DNS with superfine resolution. This study showed that there are two distinct regimes in the dissipation range: a far-dissipation range (FDR) for $k\eta > 6$ consistent with a pure exponential; and a near-dissipation range (NDR) in the vicinity of $k\eta \gtrsim 1$, where the spectrum is a pure exponential at very low $\re$ ($\gamma=1$), but  evolves into a stretched exponential with decreasing $\gamma <1$ as $\re$ increases. This analysis in \cite{kds18} was restricted to $\re \lesssim 100$, which invites the question as to whether some asymptotic high-$\re$ limit for the NDR (and hence $\gamma$) exists. 

Our goal is to assess the picture by means of a well-resolved DNS of isotropic turbulence based on highly accurate Fourier pseudo-spectral methods, going up to grids of $12288^3$ and Taylor-scale Reynolds number $\re$ ranging from 140 to 1300. The largest $\re$ here is more than an order of magnitude larger than in \cite{kds18}. We shall also interpret the findings in terms of two recent theoretical predictions; the first, resulting from ideas based on distributed chaos, predicts $\gamma=3/4$ or $2/3$ depending on a particular choice of parameters \cite{bershadskii2015}; and the second, emerging from a non-perturbative renormalization group (NPRG) approach, predicts that $\gamma=2/3$ \cite{canet17}. Both references have claimed an agreement in their respective inspections with experimental or DNS data \cite{bershadskii2015,canet17,debue18,gnova19} but, as mentioned earlier, the data were restricted to either low $\re$ or limited resolution. We assess these claims and show that there exists an intermediate bridging region between the stretched exponential and the FDR, on which we shall remark only briefly.

\paragraph*{DNS data:} The data, summarized in Table~\ref{tab:param}, are an extension of those utilized in a recent work \cite{BPBY2019}; we have also extended the runs at $\re=390$ and $650$ for longer computational times. In addition, we have performed a new run at $\re=1300$, with a small-scale resolution $k \eta = 3$ \cite{buaria_nc,buaria_wsw}. 
The totality of the data allows us to demonstrate that the behavior of the spectrum in the dissipation range, while being consistent with \cite{kds18}, is more complex at higher $\re$ than was anticipated there.

\begin{table}[h]
\centering
    \begin{tabular}{cccccc}
\hline
    $\re$   & $N^3$    & $k_{max}\eta$ & $T_E/\tau_K$ & $T_{sim}$ & $N_s$  \\
\hline
    140 & $1024^3$ & 5.82 & 16.0 & 6.5$T_E$ &  24 \\
    240 & $2048^3$ & 5.70 & 30.3 & 6.0$T_E$ &  24 \\
    390 & $4096^3$ & 5.81 & 48.4 & 2.8$T_E$ &  35 \\
    650 & $8192^3$ & 5.65 & 74.4 & 2.0$T_E$ &  40 \\
   1300 & $12288^3$ & 2.95 & 147.4 & 20$\tau_K$ &  18 \\
\hline
    \end{tabular}
\caption{Simulation parameters for the DNS runs used in the current work: the Taylor-scale Reynolds number ($\re$), the number of grid points ($N^3$), spatial resolution ($k_{max}\eta$), ratio of large-eddy turnover time ($T_E$) to Kolmogorov time scale ($\tau_K$), the simulation time ($T_{sim}$) in statistically stationary state, and the number of three-dimensional snapshots ($N_s$) used for each run to obtain the statistics.
}
\label{tab:param}
\end{table}

\begin{figure}
\begin{center}
\includegraphics[width=0.47\textwidth]{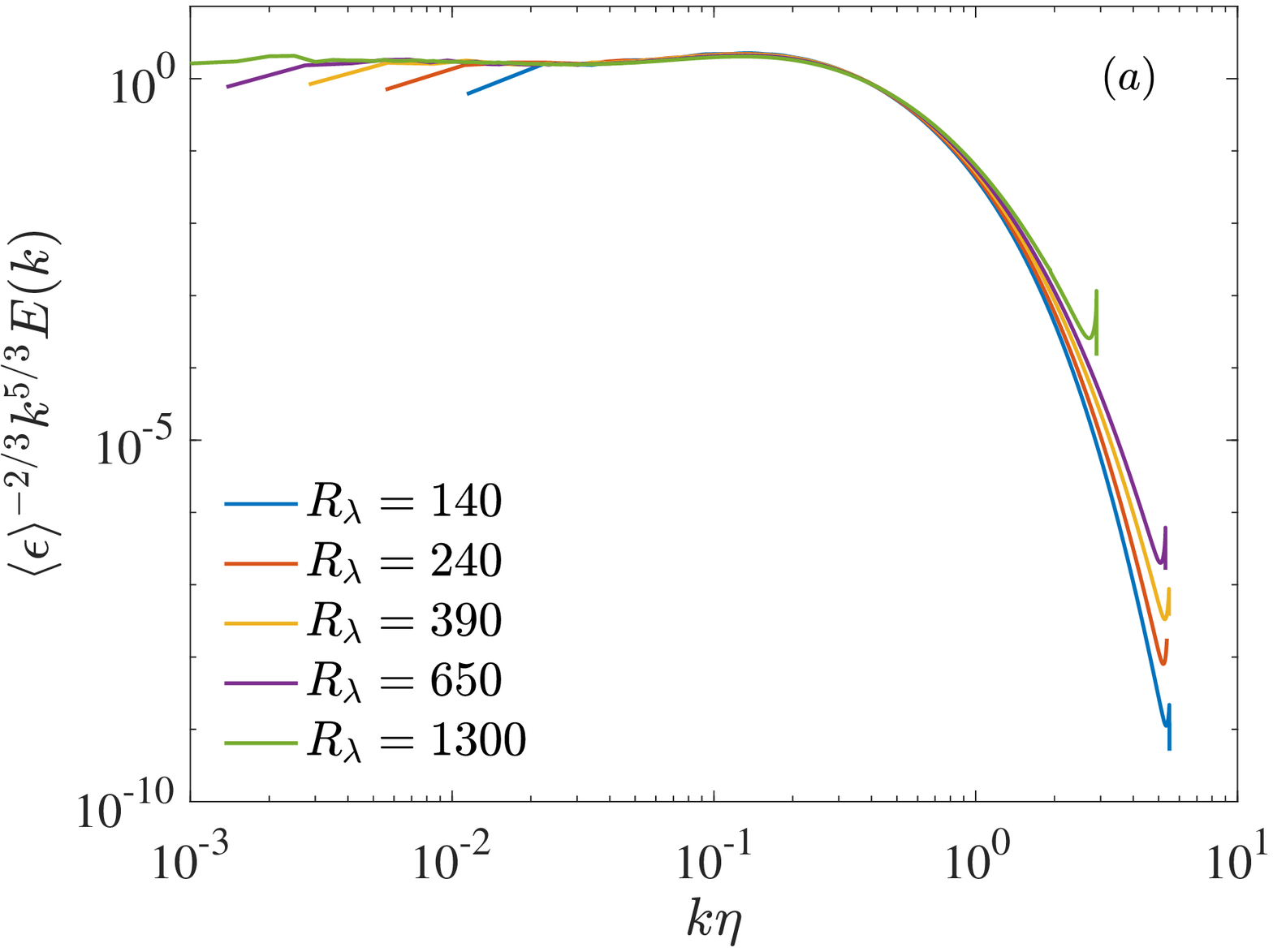} \ 
\includegraphics[width=0.45\textwidth]{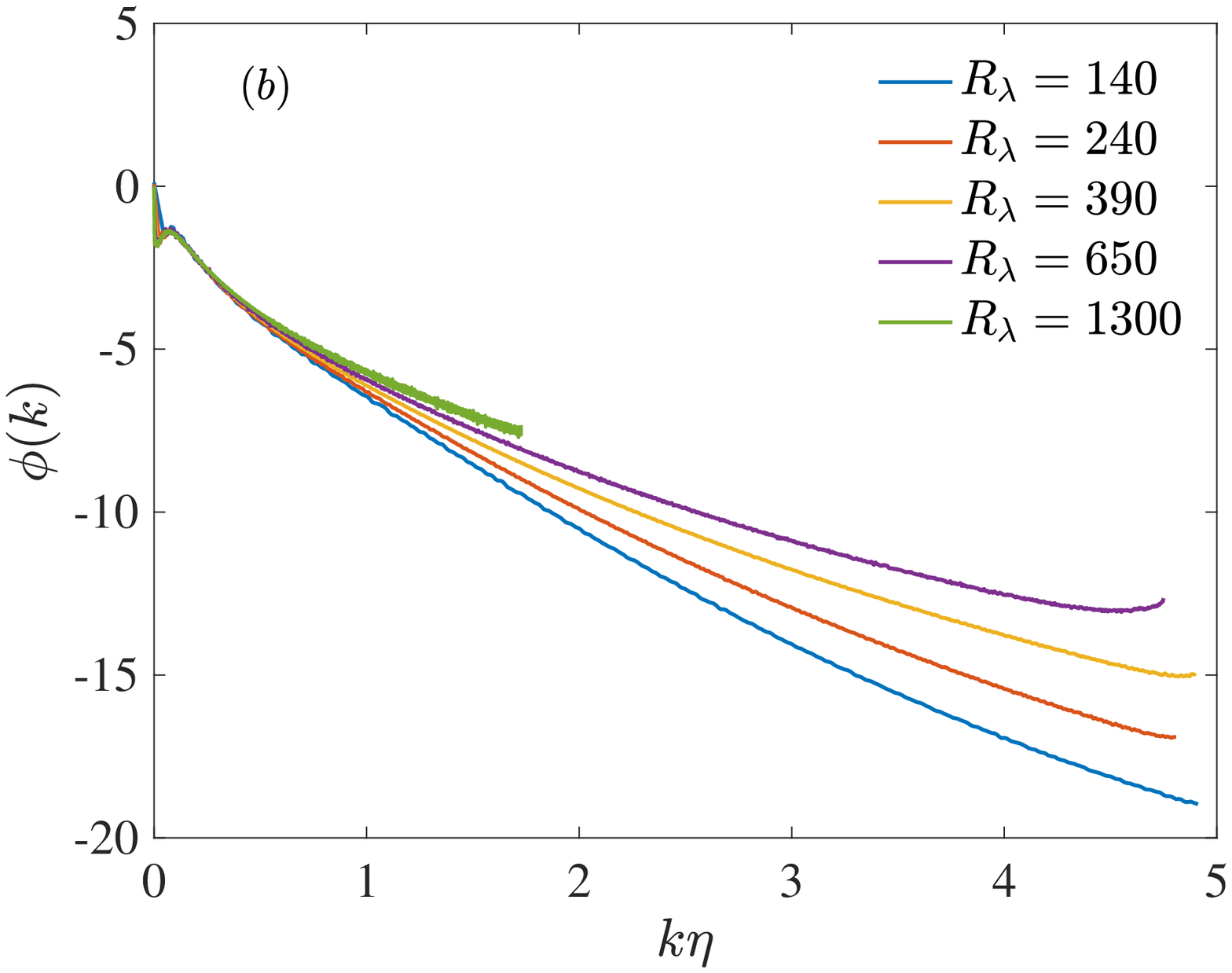}
\caption{
(a) Compensated kinetic energy spectra $E(k)$ as a function of $k\eta$ for various Taylor-scale Reynolds number $\re$. (b) The log-derivative of the energy spectra, i.e., $\phi(k) = d \log E(k) /d \log k$.  
}
\label{fig:spec}
\end{center}
\end{figure}

\paragraph*{The stretched exponential region of NDR:} In Fig.~\ref{fig:spec}a we show the compensated energy spectra for various $\re$, as a function of $k\eta$. Consistent with earlier results at lower $\re$ \cite{kds18}, a systematic enhancement in the high-wavenumber spectral density is observed with respect to $\re$. The curves clearly show that $f(k\eta)$ is not a universal function of its argument. Earlier studies such as \cite{MY.II,sv94,ishihara2005a}, which inferred a spectral collapse consistent with K41 phenomenology, were limited by technical reasons: the scatter in \cite{MY.II} was sufficiently large that possible trends could have been easily obscured; and the spectral resolution in yet others was limited to $k\eta < 1$. In order to explore the behavior further, we consider its log-derivative of Eq.~\eqref{eq:gamma}, given by
\begin{align}
\phi(k) = \frac{d \log E(k)}{d \log k} = \alpha - \beta \gamma (k\eta)^\gamma. 
\label{eq:gamma2}
\end{align}
This form allows us to isolate the stretched exponential behavior in a meaningful way.

Figure~\ref{fig:spec}b shows $\phi(k\eta)$ for various $\re$. The curves clearly suggest that the $f(k\eta)$ is non-universal and exhibits concave curvatures, confirming that $\gamma < 1$. In contrast to the results of \cite{debue18,gnova19}, Fig.~\ref{fig:spec}b at higher $\re$ shows that the energy spectra in NDR cannot be described by one single value of $\gamma$. We now undertake a more detailed analysis to extract $\gamma$ and its dependence on $\re$. We also make a preliminary note that the multifractal formalism should yield a nearly constant form for $\phi(k\eta)$ \cite{frisch91}, quite unlike the data (details are discussed later).

As noted in \cite{kds18} and other similar contexts \cite{BPBY2019}, extracting $\gamma$ through a direct curve fit of Eq.~\eqref{eq:gamma2} results in a complex nonlinear regression, which is strongly dependent on initial seeds and does not guarantee proper convergence. Hence, alternative strategies must be utilized. We adopt a modified version of the strategy utilized in \cite{kds18}. In order to evaluate $\gamma$, the authors of \cite{kds18} compensated $\phi(k)$ by $(k\eta)^\gamma$ for different $\gamma$ values, until a reasonable plateau was observed in the chosen fitting range. Furthermore, they noted that the precise value of $\alpha$ was inconsequential for the fit (because a reasonable determination scheme yields only small values with significant fluctuations), and one can set it to zero without any loss of fidelity. Consequently, Eq.~\eqref{eq:gamma2} reduces to $-\phi(k\eta) \sim \beta\gamma (k\eta)^\gamma$, and one can obtain $\gamma$ by simply fitting a power law for $-\phi(k\eta)$ in the desired range. This procedure is similar to that of \cite{kds18}, but as we will see, it has the added benefit of also identifying the appropriate ranges of power law behaviors. Other methods for extracting $\gamma$ are also possible, e.g. see \cite{debue18,gnova19}, but as described in the Appendix, they are not very robust and can lead to incorrect conclusions, especially when the $\re$ is low.

\begin{figure}
\begin{center}
\includegraphics[width=0.47\textwidth]{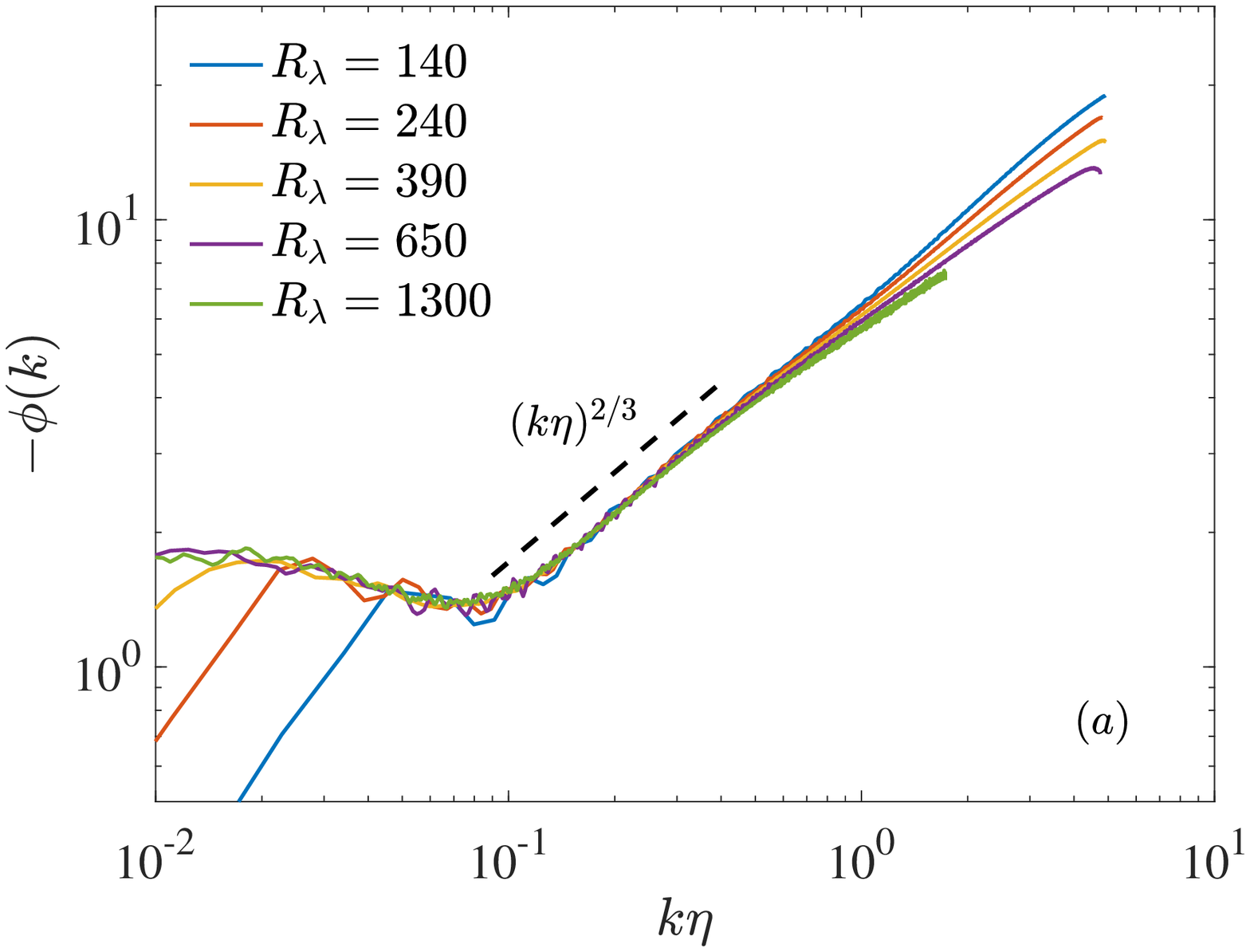}
\includegraphics[width=0.45\textwidth]{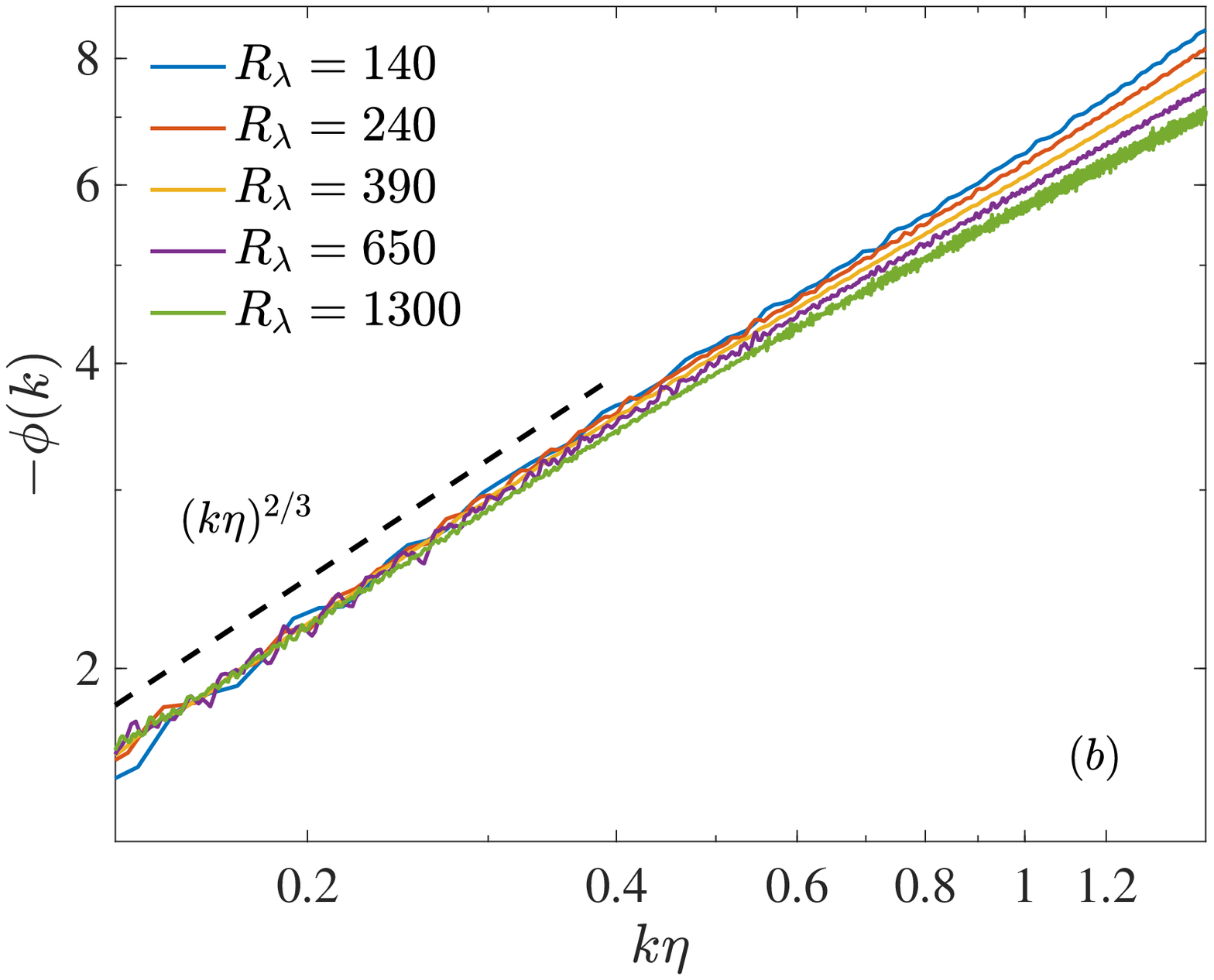}
\caption{
(a) The negative of log derivative of the energy spectra for various $\re$. (b) Zoomed in version of the same plot. The dashed blacks line in both panels represent a power law with exponent $2/3$. In the range $k\eta>1$, we observe only power laws with only a varying exponent from $0.70\pm0.03$ at $\re=140$ to $0.50\pm0.01$ at $\re=1300$. 
}
\label{fig:phi2}
\end{center}
\end{figure}

Figure \ref{fig:phi2}a shows log-log plots of $-\phi(k\eta)$ versus $k\eta$ and confirms that the log-derivative exhibits two regions of distinct power laws. (An expanded version is provided in Fig.~\ref{fig:phi2}b.) In the first, corresponding to the region immediately past the bottleneck (known to occur around $k\eta\approx 0.1$\cite{donzis2010a,kuech19}) to $k\eta \lesssim 0.5$, data for all $\re$ exhibit a spectral collapse, with the exponent ranging from $0.68\pm0.03$ for $\re=140$ to $0.67\pm0.01$ for $\re=1300$, effectively 2/3. This value of $\gamma$ is in agreement with the theoretical prediction from NPRG \cite{canet17} (though a precise wavenumber range is not obtainable from the theory). However, the analysis also predicts a strong $\re^{-3}$ dependence of the coefficient $\beta$ in the this range. Given the collapse obtained in Fig.~\ref{fig:phi2}, it follows that $\beta$ is independent of $\re$ in this range---which invites a possible refinement of the underlying theoretical arguments in \cite{canet17}. 

Another prediction in \cite{bershadskii2015} on the basis of distributed chaos yields $\gamma=3/4$, which seems to be ruled out. However, the same author provided an alternative argument that yields $\gamma=2/3$, which is consistent with the present results. Incidentally, some support for $\gamma=2/3$ was also provided in a recent work \cite{gnova19} in the range $\re=60-240$, though the fitting range included part of the wavenumber range ($0.2 \lesssim k\eta \lesssim 4$) that lies outside this range of universal fit---and thus produced the considerable error bar. Our results show that the prediction from NPRG is valid only in a small region of NDR and the behavior in the remainder of NDR ($k\eta>1$) is quite different, as shown next.

\begin{figure}
\begin{center}
\includegraphics[width=0.47\textwidth]{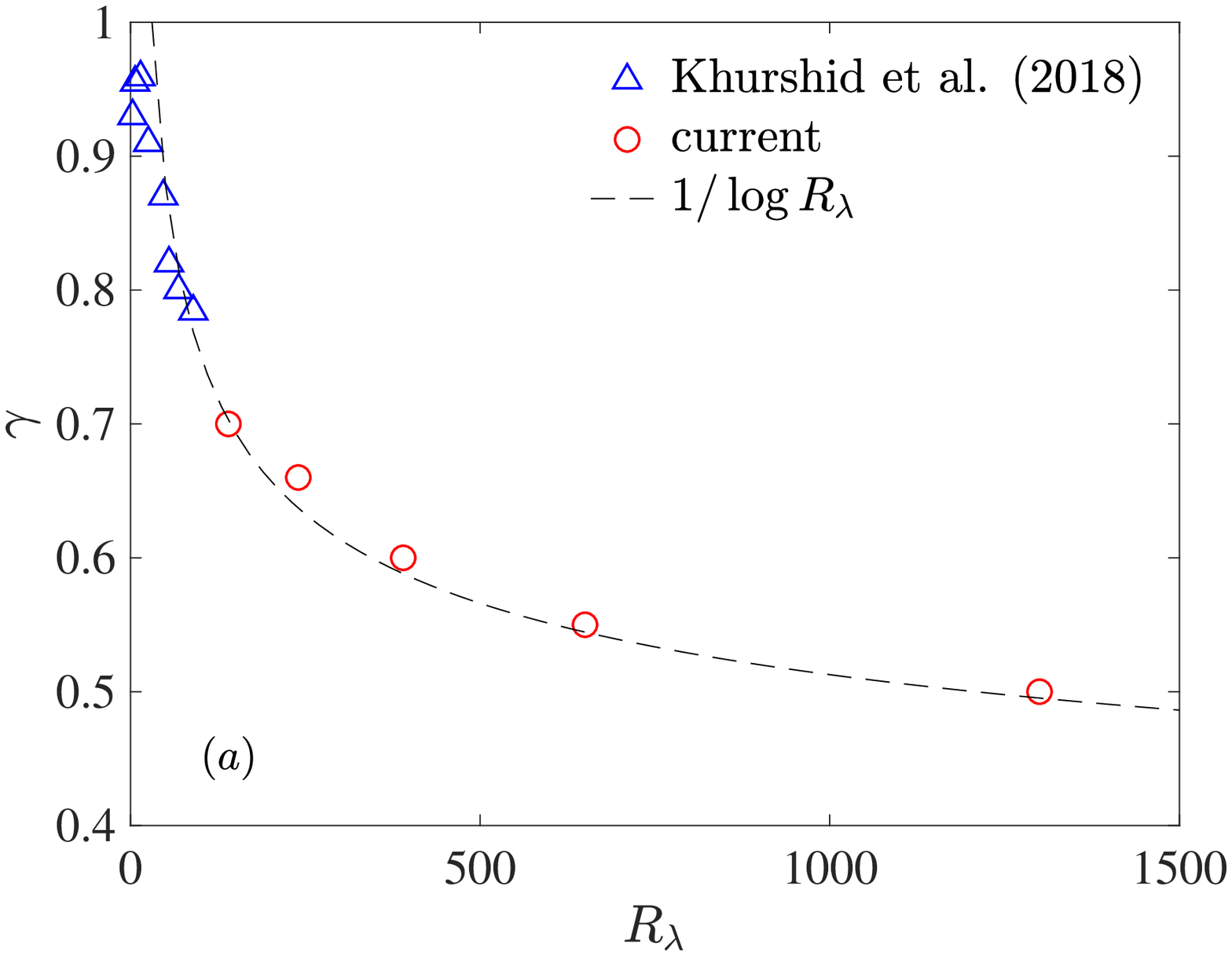}
\includegraphics[width=0.45\textwidth]{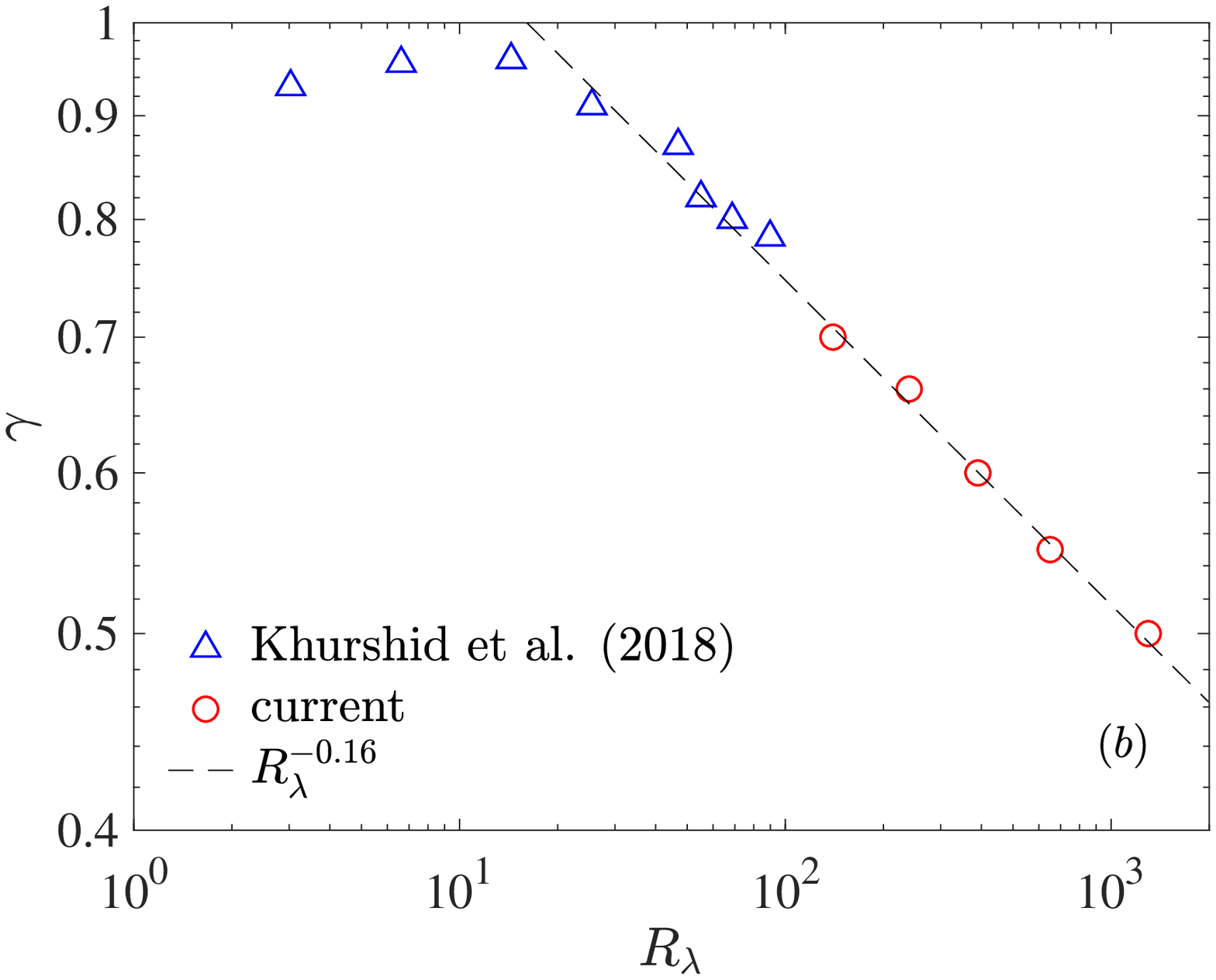}
\caption{The exponent $\gamma$ as a function of $\re$ on (a) linear scales, and (b) log-scales. The (blue) triangles correspond to data of \cite{kds18}. The dashed lines correspond to fits shown in the legend.
}
\label{fig:gamma}
\end{center}
\end{figure}

\paragraph*{The remainder of NDR:} From Fig.~\ref{fig:phi2}, the second region where power laws can be fitted is the range $1<k\eta<4$, which is similar to that utilized in \cite{kds18,gnova19}. Note that the range is slightly smaller for $\re=1300$, since the data do not go beyond $k\eta=3$. It is clear that no single value of $\gamma$ is adequate to describe the entire NDR, consistent with the results of \cite{kds18} at lower $\re$. We have plotted the current data in Fig.~\ref{fig:gamma}a together with the data for $\re \le 100$ from \cite{kds18}. Evidently $\gamma$ continues to decrease for the $\re$ range considered here, with a plausible fit that is logarithmic (on which we comment later). Alternatively, Fig.~\ref{fig:gamma}b shows an equally plausible weak power law with $\gamma \sim \re^{-0.16}$, for $\re > 20$, say. 

Both these dependencies are similar to how the bottleneck flattens with the Reynolds number \cite{donzis2010a}. In fact, it seems reasonable that the increase in spectral density (with $\re$) in the dissipation range is connected to a decrease in the bottleneck region. Physically, the bottleneck is thought to develop due to inadequate `thermalization' of the energy transferred from inertial to dissipation scales, leading to a pileup at their crossover \cite{frisch08}. However, with increasing $\re$ and the scale-range, the energy transfer across the scales is better facilitated, leading to the diminution of the bottleneck and simultaneous rise in spectral density in NDR. Recent experimental results \cite{kuech19} have confirmed the decay of the bottleneck even up to $\re\approx4000$. Based on this behavior, we may infer that the exponent in this part of NDR will likely continue to decrease at least up to $\re=4000$; however, if the trend in Fig.~\ref{fig:gamma}b persists for higher $\re$, it is clear that the asymptotic value will be zero, achieved probably at extremely high Reynolds numbers.

It is useful to note that the multifractal (MF) analysis of \cite{frisch91} predicts a log-dependence of the exponent $\gamma$. Their analysis predicts the spectrum to have the form $\log E(k\eta)/\log \re = F(\theta)$, where $\theta=\log k\eta/\log \re$, and $F$ is supposedly universal. Taking the log-derivative gives $\phi(k\eta) =  F'(\theta)$, which can be reconciled with Eq.~\eqref{eq:gamma2} (and hence the stretched exponential behavior) if $\gamma$ scales as $1/\log \re$ (since $\theta = \log (k\eta)^{1/\log\re}$, which must match the $(k\eta)^\gamma$ behavior). The fit shown in Fig.~\ref{fig:gamma}a is indeed consistent with this expectation. However, it should also be noted that the MF analysis of \cite{frisch91} also predicts the precise functional form of $F(\theta)$ in the NDR, which is similar to a power law dependence of the spectra (since $F(\theta)$ is an algebraic
function of $\theta$). This prediction is clearly not consistent with the current data (see Figure~\ref{fig:spec}b). Also, as noted in \cite{kds18}, the MF prediction does not appear to work for the spectra in the FDR ($k\eta\gg1$), to which we will return later.  One likely reason for the disagreement is that the arguments presented in \cite{frisch91} are valid in the large Reynolds numbers. This would be consistent with recent work of \cite{BPBY2019}, which suggested that the assumptions built in to the MF model can be realized only at astronomically high $\re$ (that are impossible to simulate, even without the fine resolution used here). Nevertheless, one has to leave open the question of whether the trend observed here for $\gamma$ holds up for much higher $\re$.

An alternative application of the MF model is based on the
extension of approximate parameterizations for the second order 
structure functions, aimed at characterizing the transition 
between viscous and inertial-range scalings
\cite{meneveau96,stolo93,chevillard06,arneodo08,bos12}. 
The energy spectrum can be indirectly obtained by appropriately taking the Fourier transform of the second order structure function. However, as noted in \cite{meneveau96} and references therein, such parameterizations are not necessarily unique.
Moreover, they also do not explicitly predict a stretched exponential 
function as considered here. 
Nevertheless, it would still be instructive to utilize the 
current high-resolution DNS data to test these
approaches by directly investigating the structure 
functions instead of the energy spectra, which we leave
for future work.


Finally, we note that $\beta$ in Eq.~\eqref{eq:gamma2} is also a parameter of the stretched exponential fit. Given how the NDR beyond $k\eta>1$ departs systematically from the universal regime with $\gamma=2/3$, it follows that the product $\beta\gamma$, the coefficient that appears in Eq.~\eqref{eq:gamma2}, will emerge as independent of $\re$ (since these power law fits can be thought to have a common origin with different slopes). This implies that $\beta \sim 1/\gamma$. While this inference is consistent with the observation in \cite{kds18}, the precise value of the product $\beta\gamma$ is strongly dependent on the exact fitting range and hence not very useful. 

\paragraph*{The far dissipation region:} As far as we know, only the authors of \cite{kds18} were able to adequately resolve the range $k\eta > 6$ (FDR). They could do it because of the comparatively low $\re$ of their simulations. Their conclusion is that the spectral shape in FDR is exponential, consistent with analytical arguments \cite{kraichnan1959a, frisch1981a, sreenivasan1985a,foias1990a,sirovich1994a} that require viscosity to regularize the velocity field at very small scales. Thus, it appears reasonable to expect that the spectrum in the FDR would be a pure exponential. It has not been possible for us to have the same resolution and also extend to the $\re$ values attained in this paper. Thus, we leave open the possibility that a pure exponential occurs for wavenumbers higher than $k\eta > 6$ even at very high $\re$. It is not lost on us that the increasing demands on resolution could be hinting something important at the analytic structure of the Navier-Stokes equations.

\paragraph*{Concluding remarks:} We have analyzed the dissipation range behavior of the energy spectra obtained from very well resolved DNS of isotropic turbulence at Taylor-scale Reynolds numbers that are an order of magnitude higher than in earlier studies. In the process, we have extended the work of \cite{kds18} and also undertaken the verification of various theoretical predictions. Our results indicate that the behavior of the spectra in NDR is more complex than previously realized. In a limited range of NDR, $0.15 \le k\eta \le 0.5$, our results show a universal stretched exponential fit to the spectra, of the form $\exp(-k^{2/3})$. This result matches the theoretical prediction from NPRG, but the anticipated range of validity is much smaller than that asserted in recent works \cite{canet17,gnova19}. It is also consistent with one version of the distributed chaos \cite{bershadskii2015}. In the FDR, one can anticipate a pure exponential predicted from analyticity arguments \cite{kraichnan1959a}. However, the behavior of the spectra in the near dissipation range $1< k\eta < 4 $ still remains an open question. While the spectra are consistent with stretched exponential behavior in this range, our data show that the exponent decreases with the Reynolds number, without a tendency to asymptote. Evidently, further theoretical developments are necessary to
explore this behavior with confidence.


\section*{Acknowledgments}

We thank Alain Pumir, Diego Donzis, P. K. Yeung and Sualeh Khurshid for their comments on the draft and sustained collaboration over the years. We gratefully acknowledge the Gauss Centre for Supercomputing e.V. (www.gauss-centre.eu) for providing computing time on the GCS supercomputers JUQUEEN and JUWELS at J\"ulich Supercomputing Centre (JSC), where the simulations were performed. This work was also partly supported by supercomputing resources under the Blue Waters sustained petascale computing project, which was supported by the NSF (awards OCI-5725070 and ACI-1238993) and the State of Illinois.

\appendix

\section{Robust determination of the exponent in stretched exponential curve fit}

\begin{figure}[h]
\begin{center}
\includegraphics[width=0.47\textwidth]{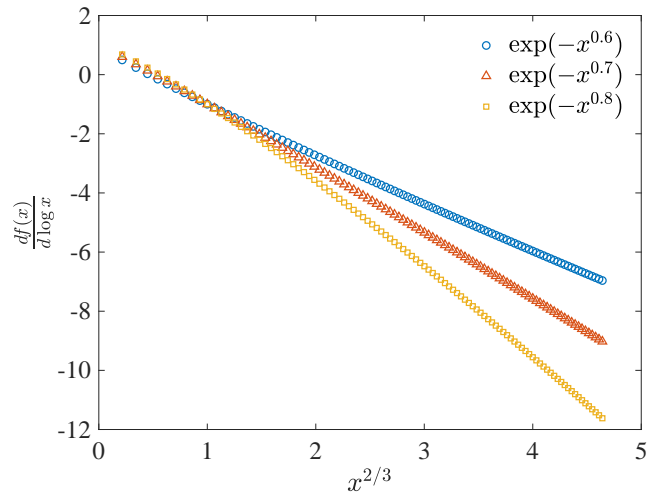}
\caption{
Log-derivative of various stretched exponential functions $f(x)$, plotted versus $x^{2/3}$. All curves exhibit a visually perceptible range where they look like a straight line. Similar behavior is observed if some other power of $x$ is used instead of $2/3$. 
}
\label{fig:error}
\end{center}
\end{figure}

In determining the exponent $\gamma$ in Eq.~\eqref{eq:gamma2} from experimental or numerical data, a few different methods can be employed (other than determining the power-law exponent as done in the current work). One method, also utilized in recent works \cite{bershadskii2015,canet17,debue18,gnova19}, is to plot the log-derivative $\phi(k\eta)$ vs. $(k\eta)^\gamma$ for a choice of $\gamma$ and thereafter compare the curve with a straight line. However, we note that this method relies heavily on a visual comparison, rather than an explicit curve fit, and is inherently error prone \cite{BSY.2015}. For instance, in Fig.~\ref{fig:error} we simply plot the log-derivative of various exponential functions $f(x)$ as a function of $x^{2/3}$. As is evident, all curves can be erroneously matched with a straight line on the basis of a simple visual inspection, leading to the incorrect conclusion such as the exponent being $2/3$ in a larger range.

Another method is to directly determine the log-derivative of $-\phi(k\eta)$ with respect to $k\eta$, which in principle allows for a direct evaluation of $\gamma$ from the resulting local slopes plot, without any explicit curve fit. However, if the parameter $\alpha$ is not set to zero, one needs to evaluate three successive log-derivatives, as was done in \cite{gnova19} at significantly lower $\re$ than here. We did not find this method to be reliable for our data, since calculating three log-derivatives leads to substantial numerical noise, making it nearly impossible to meaningfully extract the exponent. It is possible that this effect is less pronounced at low $\re$ \cite{gnova19}, but does not work at high $\re$ investigated here. Finally, we note that in the method employed in \cite{kds18} $\phi(k\eta)$ is compensated by $(k\eta)^\gamma$ until a reasonable plateau is obtained. While this indeed results in a reasonable fit, it requires an advance knowledge of the fitting range (which perhaps is the reason why the $2/3$ region was overlooked in that work).

%

\end{document}